\documentclass[preprint]{aastex}

\newcommand{\sech}{\mbox{sech}}
\newcommand{\sign}{\mbox{sign}}
\newcommand{\DM}{\mbox{DM}}

\newcommand{\pc}{\mbox{~pc}}
\newcommand{\kpc}{\mbox{~kpc}}
\newcommand{\cm}{\mbox{~cm}}

\slugcomment{Submitted to Astronomical Journal}

\shorttitle{Distribution of Galactic Free Electrons}
\shortauthors{}

\begin{document}


\title{A Re-examination of the Distribution of Galactic Free Electrons}


\author{Gilberto C. G\'omez\altaffilmark{1,2},
        Robert A. Benjamin\altaffilmark{2}, and
        Donald P. Cox\altaffilmark{2}}
\affil{
       University of Wisconsin - Madison \\
\altaffilmark{1}Department of Astronomy, 475 N. Charter st.,
                Madison WI, 53706 \\
\altaffilmark{2}Department of Physics, 1150 University Avenue,
                Madison, WI 53706; \\
                gomez@wisp.physics.wisc.edu, benjamin@wisp.physics.wisc.edu,
                cox@wisp.physics.wisc.edu}




\begin{abstract}

We present a list of 109 pulsars with independent distance information
compiled from the literature. Since the compilation of Frail \& Weisberg,
there are 35 pulsars with new distance estimate and 25 pulsars for which
the distance or distance uncertainty have been revised.  We used this
data to fit a smooth, axisymmetric, two disk model of the distribution
of galactic electrons.  The two exponential model components have mean local
midplane densities at the solar circle of
$2.03 \times 10^{-2}
\cm^{-3}$ and $0.71
\times 10^{-2}
\cm^{-3}$, and scale heights of 1.07 and 0.053 kpc. The
thick component shows very little radial variation, while the second has
a radial scale length of only a few kiloparsecs.  We also examined a model
which varies as
$\sech^2(x)$, rather than $\exp(-x)$, in both the radial and vertical
direction.  We prefer this model with no midplane cusp, but find that
the fit parameters essentially describe the same electron
distribution.  The distances predicted by this distribution have a
similar scatter as the more complex model of Taylor \& Cordes.
We examine the pulsars that deviate strongly from this model.
There are two regions of enhanced dispersion measure, one of which
correlates well with the Sagittarius-Carina spiral arm. We find that the
scatter of the observed dispersion measure from the model is not
fit well by either a normal or log-normal distribution of lump sizes, but
may be caused instead by the uncertainties in the distances.  

\end{abstract}



\section{Introduction}

One of the most important discoveries in the study of the interstellar
medium is the realization that the warm ionized medium (WIM) is a major
component of our galaxy; it has a thick distribution and it is not
localized around ionization sources. Reynolds (1989) used the dispersion
measures ($\DM= \int n_e dl$) towards pulsars with known distance to
measure the scale height of the WIM in the Milky Way, and showed that it
is much larger than that of the bulk of the neutral hydrogen.

If the distance to a pulsar is known, this can be used with its DM to
constrain models for the spatial distribution of the free electrons. The
most popular model is that derived by Taylor \& Cordes (1993, TC), which
also used the observed scattering measures to a different set of pulsars
to refine the parameters of the model.
Their most important contribution was the addition of non-axisymmetric elements,
i.e., spiral arms defined by the locations of HII regions (Georgelin \& Georgelin
1976).  Their main justification is the observed asymmetry in the DM vs.
galactic longitude plots. They also incorporated the unusually high DM
observed towards the Gum Nebula.

Models of this kind are used frequently to determine distances to
pulsars. TC claim that their model yields distances
accurate  to 25\%. But, since its publication, the set of pulsars with
independent distance measurement has increased, some distances have
been revised, and pulsars with forbidden DM (higher than the
asymptotic value predicted by TC) have been observed.  In addition,
observations of the the angular broadening of radio sources has been
used to constrain the electron density in the Galactic center (Lazio
et al. 1999; Lazio \& Cordes 1998a,b) and the scalelength of the
distribution in the anti-center direction (Lazio \& Cordes
1998c,d). Finally, the recent completion of the Wisconsin H$\alpha$
Mapper survey of diffuse galactic H$\alpha$ emission with one-degree
angular resolution and $\sim 10~ {\rm km~s^{-1}}$ velocity
resolution (Reynolds 1998; Haffner 2000) will allow the development of
more complex models. These observations will allow for a reassessment
of the location of galactic spiral arms (Georgelin \& Georgelin 1976;
Ruseil et al. 1998; Georgelin et al. 2000), as well as the discovery and
placement of large angular-scale HII regions, such as the Gum
Nebula.

In this work, we present an updated list of pulsars of known
distance. We then use this data to constrain a new axisymmetric model
for the free electron distribution, and show how the Taylor-Cordes
model and the new axisymmetric model fare in predicting distances to
the pulsars. We also consider to what degree the available data
constrain the lumpiness of the warm ionized medium. Incorporation of
non-axisymmetric effects, such as the galactic spiral arms and
individual nebulae can subsequently be incorporated using the WHAM
data and more recent radio recombination line surveys of HII regions.

\section{The Pulsar Data Set}

A list of 109 pulsars with distance information was gathered from a
number of sources and are compiled in Table 1, and presented in order of
increasing distance.
Of this list, four are
in the Large or Small Magellanic clouds. Of the remainder, 76 have
both upper and lower distance limits; 20 have only lower limits, and 9
have only upper limits. This dataset is $\sim$ 50\% larger than the
data used by TC. Of the 109 pulsars, there are 35 new
distance determinations since the compilation of Frail \& Weisberg
(1990, FW90), which provided the bulk of the measurements used in the TC
model, 25 objects in which there have been revisions in
either the distance or distance uncertainty of the pulsar, and 49 objects whose distance estimates remained unchanged.

Distance estimates come from a variety of methods, which we briefly
summarize here.

{\it Kinematic distances (68 pulsars):} The majority of pulsar
distance measurements come from the combination of 21 cm absorption
combined with an axisymmetric, kinematical model for galactic rotation
(Fich, Blitz, \& Stark 1989). FW90 re-evaluated all the distance
measurements up to that time using this model (with corrections for
pulsars towards the Perseus arm), and a uniform set of criteria for
converting absorption velocities to distance. These criteria have been
adhered to in subsequent work. Probably the largest source of
systematic error is due to the  non-circular 
``streaming'' motions in the vicinity of spiral arms.

{\it Association with globular clusters (17 pulsars):} The next most
common distance determination method come from association of a pulsar
with a globular clusters of known distance. Table 1 only lists one
pulsar per globular cluster; when more than
one pulsar is known, the variation in dispersion measure is
small. Since the compilation of FW90, the distances to globular
clusters have been considerably refined due to improved
color-magnitude diagrams and shifts in the assumptions about the
luminosity of RR Lyrae stars. As a result, some distances estimates
have been revised by more than a factor of two since FW90 (Harris
1996, with online updates at
http://physun.mcmaster.ca/$\sim$harris/mwgc.dat).  The uncertainty in the
distance modulus of these clusters was assumed to be $\sigma=0.1+0.4
E_{B-V}$ magnitudes. More heavily reddened clusters have poorer data
since they have greater problems with field contamination and crowding
(Harris 1999).

{\it Association with supernova remnants (10 pulsars):} There have
been numerous suggested associations between pulsars and supernova
remnants (Lorrimer 1998; Gaensler \& Johnston 1995; Frail, Goss, \&
Whiteoak 1994; Kaspi et al. 1996) . However, such associations are hard
to prove, since they depend upon expectations for supernova remnant
lifetimes, pulsar ages, and transverse velocities. In this
compilation, we use the associations judged by Lorrimer, Lyne, and
Camilo (1998) to be the ``most likely'' pulsar-supernova remnant
pairs.  The only other pulsar/SNR associations added were B1800-21
with G8.7-0.1 (Finley \& \"{O}gelman 1994) and B1758-23 with W28
(Frail, Kulkarni, \& Vasisht 1993). Both of those have independent kinematic
distances which support the association. 

{\it Trigonometric parallax (8 pulsars):} Potentially the most
reliable distances come from interferometric measurements of annual
parallax. However, there are several practical difficulties arising
from ionospheric effects and a scarcity of nearby calibrators for
positions. Improvements in the techniques have led to changes in the
published distances by more than a factor of two. The distance
estimate for B0950+08 increased from 130 pc (Gwinn et al. 1986) to 280
pc (Brisken et al. 2000), while the distance estimate for B1919+06
decreased from 3.3 kpc (Fomalont et al. 1999) to 1.2 kpc (Chatterjee et
al 2000). These changes were much larger than the stated uncertainties
in the measurements. Accurate estimates are vital if pulsars are to be
used as probes of the structure of the local interstellar medium.

{\it Association with other galaxies (4 pulsars):} Four pulsars have
been associated with the Magellanic Clouds, three in the LMC, one in
the SMC. These pulsars are valuable in constraining the electron
density in the Galactic halo. However, an unknown fraction of the
dispersion measure must arise in the host galaxy, so their utility in
constraining the Galactic free electron column density is compromised.

{\it Timing parallax (5 pulsars):} Distances to millisecond pulsars
have also been estimated using variations in arrival time of the
pulses. There is a annual change in the pulse arrival time whose
magnitude is given by $\Delta t=r^{2}cos^{2} \theta/(2cd)$ where r is
the Earth-Sun distance, $\theta$ is the angle between the line of
sight and the ecliptic plane, and $d$ is the distance (Ryba \& Taylor
1991). This variation is $\Delta t=1.2 \mu$ sec for $d=$ 1 kpc. This
level of timing accuracy has been reached for only a few
pulsars.

{\it Period derivative distances (2 pulsars):} Bell \& Bailes (1996)
have shown that in many cases, the observed orbital period derivative
of binary pulsars is dominated by term of the form
$\dot{P}_{b}/P_{b}=v^{2}/(cd)$. If one uses the predictions of general
relativity to derive the intrinsic period derivative, knowledge of the
proper motion of the pulsar then allows for an accurate estimate of
the distance. This method has only been applied to two pulsars to
date.

{\it Spectroscopic parallax of binary companion (1 pulsar):} There is
one case in which the binary companion of a pulsar is a $\sim$ 10
$M_{\sun}$ Be star (Johnston et al. 1994). In this case, spectroscopic
parallax was used to estimate the distance.

{\it X-ray luminosity distance (1 pulsar):} There is one distance
estimate for B0656+14 based upon the identification of the X-ray
counterpart together with a model of thermal X-ray emission from
the neutron star (Golden \& Shearer 1999). As will be seen, this pulsar
ends up being an outlier in our model. As a result, we are not
convinced that this method is reliable.

{\it Trigonometric parallax of optical counterpart (0 pulsars):} If
the optical counterpart of a pulsar can be identified, then ground
based or Hubble Space Telescope observations could yield a parallax
estimate. This technique has been used to determine the distance to
the neutron star Geminga (Caraveo et al. 1996). However, we have not
included Geminga in our list because it is unclear if it has a
reliable radio signal. The search for optical counterparts of
pulsars has been relatively unsuccessful to date (Caraveo
2000). Still, we think this method holds some promise, particular for
pulsars with the very lowest dispersion measures like J0108-1431 which
has $DM=1.83~{\rm cm^{-3}~pc}$ (Tauris et al. 1994).

{\it Scattering screen distance (0 pulsars):} It has been suggested
that the transverse velocity of a pulsar derived using models of
interstellar scintillation can be combined with measurements of proper
motion to constrain the distance to the pulsar (Gupta 1995; Deshpande
\& Ramachandran 1998; Cordes \& Rickett 1998). Application of this
model requires a knowledge of the distribution of electron density and
scattering properties along the line of sight, and as a result is
principally useful for pulsars which lie behind HII regions of known
distance or pulsars well above the disk of the Galaxy.

{\it Cross-checks (7 pulsars):} There are six pulsars for which two 
independent methods have been applied for distance determination.  In
each case, the distances estimate agree within the stated errors,
although in two cases the agreement is marginal. Such checks are
important since they test the reliability of the individual methods. We
summarize these results here. {\bf B1929+10:} This pulsar has three
discrepant measures for trigonometric parallax, $\pi=21.5
\pm 0.3$ mas (Salter, Lyne, \& Anderson 1979), $\pi < 4$ mas (Backer
\& Sramek 1982), and $\pi=5.0 \pm 1.5$ mas (Campbell 1995). The kinematic
distance is $d<1.6$ kpc (Weisberg, Rankin, \& Boriakoff 1987). We have
adopted the most recent parallax distance, which is consistent with the
kinematic distance. {\bf B0833-45:} The distance to the Vela SNR is given
as $d=250 \pm 30$ pc (Cha, Sembach, \& Danks 1999), while recent VLBI
parallax gives $d=316^{+37}_{-29}$ pc (Legge 2000). While these
uncertainties do not overlap, the uncertainties in stellar distances may
be slightly underestimated. {\bf B1855+09:} Timing parallax distance to
this pulsar was given as
$d=0.83^{+0.66}_{-0.24}$ pc (Ryba \& Taylor 1991), later refined to
$d=0.91^{+0.34}_{-0.20}$ pc (Kaspi, Taylor, \& Ryba 1994). This agrees
marginally well with the kinematic distance limits $d_{lower}=1.6 \pm
0.5$ to $d_{upper}=2.0 \pm 0.4$ (Kulkarni, Djorgovski, \& Klemola 1991).
{\bf B1800-21:} The kinematic distance limits to this pulsar are
$d_{lower}=4.0 \pm 0.6$ kpc and $d_{upper}=4.9 \pm 0.3$ kpc which agree
with the kinematic distance to the SNR G8.7-0.1, also established
kinematically (Finley \& \"{O}gelman 1994). {\bf B1758-23:} The kinematic
distance limits to this pulsar are
$d_{lower}=3.5 \pm 0.9$ kpc and $d_{upper}=6.9 \pm 0.1$ kpc which agree
with the kinematic distance to W28, also established kinematically
(Frail, Kulkarni, \& Vasisht 1994). {\bf B1937-21:} Kinematic distance
limits are $d_{lower}=4.6 \pm 1.9$ kpc and
$d_{upper}=14.8 \pm 0.9$ kpc (Heiles et al. 1983), which agrees with the 
timing parallax distance of $d > 3.6$ kpc (Kaspi, Taylor, \& Ryba 1994).
{\bf B1534+12:} The period derivative distance to this binary pulsar is
$d=1.08 \pm 0.15$ kpc, which is consistent with the timing parallax limit
of $d>0.67$ kpc (Stairs et al. 1999)

Figure 1 shows the spatial distribution of the pulsars in our
sample with both upper and lower distance limits projected onto the
Galactic plane, while Figure 2 shows a plot of their dispersion measure as a 
function of galactic longitude.  Note that in Figure 2, there is no clear
evidence for the asymmetry in maximum dispersion measure around
$l=0^{\circ}$, which is present in the complete set of pulsars including 
those of unknown distance.

\begin{figure}
\plotone{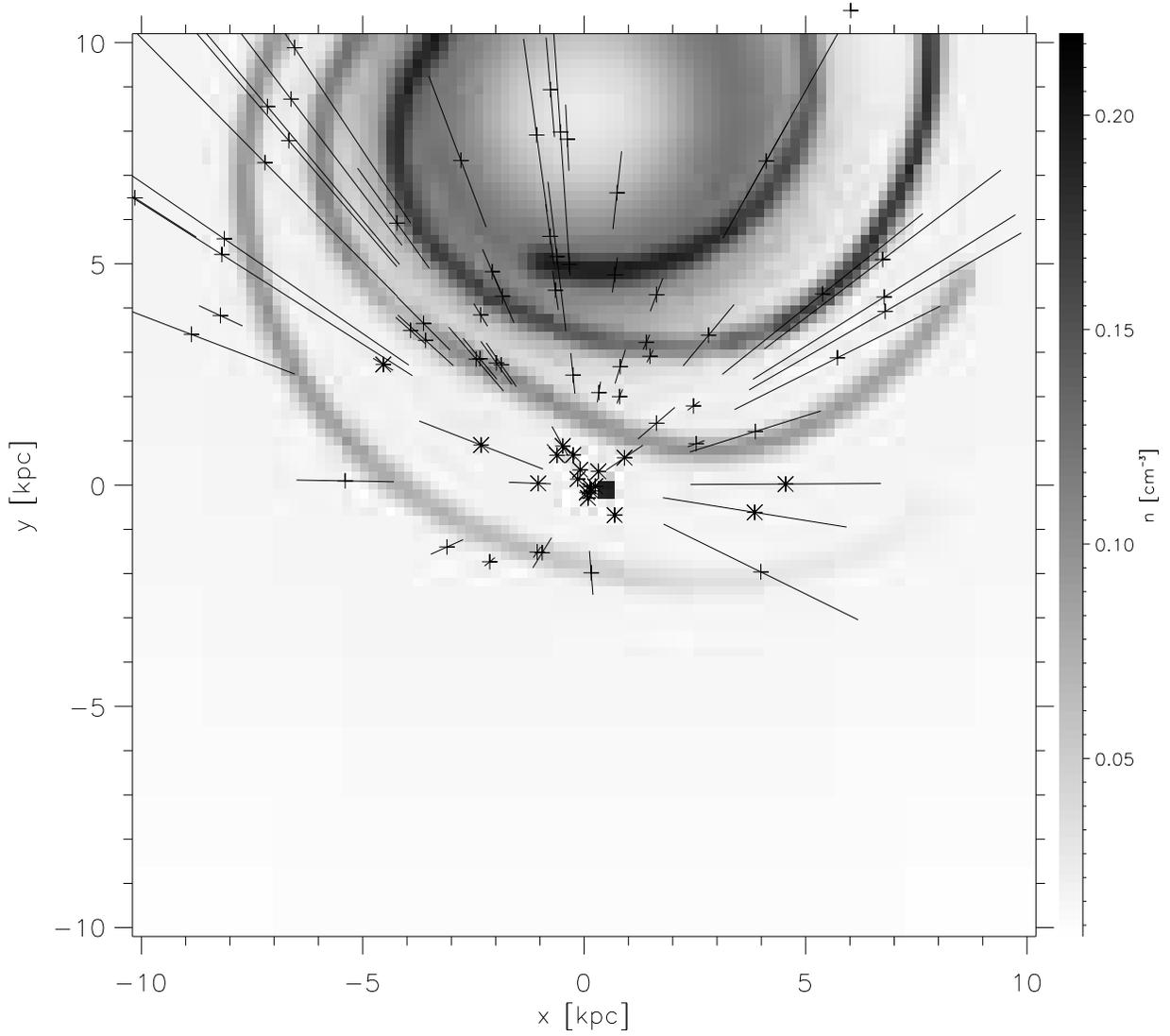}
\caption{Projection of the position of the pulsars
onto the galactic plane. Uncertainties in the distance from the Sun are
also shown. The location and density of the spiral arms,
central annulus and Gum Nebula
in the Taylor-Cordes model are noted in grey-scale. Pulsars marked
with a star are the ones considered in the ``interarm direction'' as
seen from the Sun, although this neglects the potential contribution
of the Local Arm. }
\end{figure}

\begin{figure}
\plotone{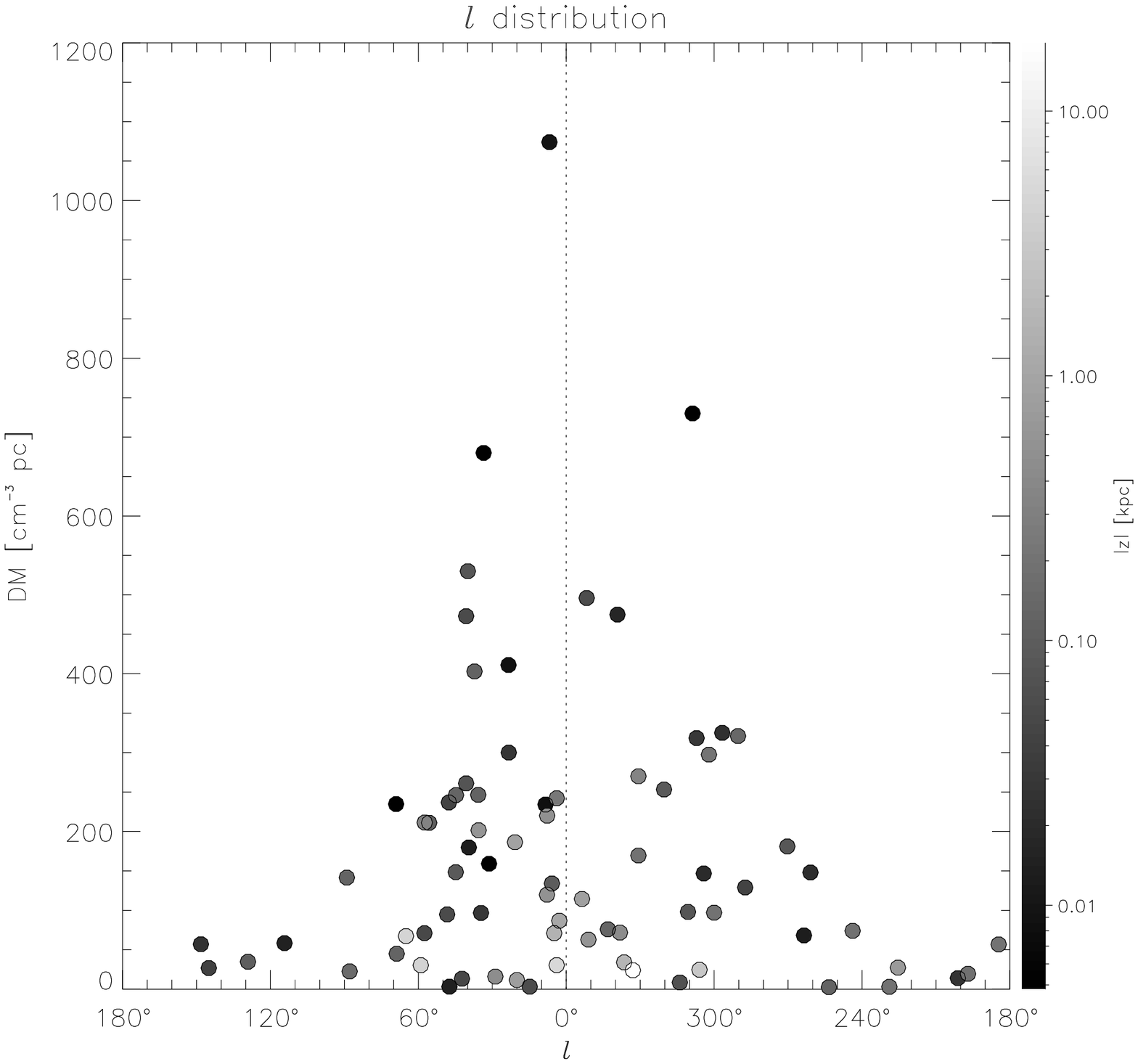}
\caption{Distribution of the pulsar dispersion
measure with galactic longitude. Grayscale indicates their distance
from the galactic midplane. In this data set, there is no clear
evidence for an asymmetrical distribution with galactic longitude. }
\end{figure}

\section{Fitting an axisymmetric model}

Using this data, we fit a two component model of the galactic disc to the pulsar
data. The model has the form,

$$n_e(r,z) = n_0 \frac{f(r/r_0)}{f(r_\Sun/r_0)} f(z/z_0) +
             n_1 \frac{f(r/r_1)}{f(r_\Sun/r_1)} f(z/z_1), $$

\noindent
where $f(x)$ is either $= \exp(-x)$ or $= \sech^2(x)$, and $r_\Sun =
8.5 \kpc$ is the galactocentric distance of the Sun.  The fit was
achieved through a variant of the $\chi^2$ method: we defined the
error-of-the-fit $\Delta$ as,

$$ \Delta = \frac{1}{n-\nu} \sum \frac{\log^2(\DM_{data}/\DM_{model})}
  {\sigma^2+\sigma_A^2}, $$

\noindent
where $n$ is the number of pulsars (76) with both upper and lower 
distance limits, $\nu$ is the number of free
parameters in the model (6), $\DM_{data}$ are the observed DM's,
$\DM_{model}$ is the modeled DM's, obtained by integrating the model
through the line of sight to each pulsar position, $\sigma = 0.5
\log(D_{max}/D_{min})$,$D_{max}$ and $D_{min}$ are the one sigma
distance brackets, $\sigma_A = 0.5 \log[(1+A)/(1-A)]$ and $A$ is a
noise parameter. The form of this extra term comes from assuming that
there is extra error proportional to the dispersion measure, i.e,

$$\DM_A = \DM_{model} (1 \pm A) $$
$$\sigma_A = 0.5 \log(\DM_{A+}/\DM_{A-}). $$

\noindent
Most of the distances to the pulsars we used (41 out of 76) are
determined by assuming a kinematic model for the galactic rotation and
comparing it to the 21 cm absorption observed towards the pulsar. For
these pulsars, we define the distance to be halfway between the
minimum and maximum limits. For these pulsars, there is a uniform
probability for the location of the pulsar between the distance
brackets, as opposed to the distances obtained by parallaxes, for example,
which have a gaussian probability distribution for the distance around
a preferred value. Therefore, the kinematic distances have an extra
$1/\sqrt{3}$ factor in the corresponding $\sigma$.

An annealing procedure was used to get the best fit for the
$n_{0,1},r_{0,1}$ and $z_{0,1}$ parameters with $A=0$.  Then, $A$ was
adjusted to get $\Delta = 1$ and a new fit was obtained.  The procedure
was repeated until convergence was achieved. Outlier pulsars were spotted
by a procedure described below, and those common to both functional forms
were taken out of the sample.  Then, the procedure was repeated and the
new fit is the one considered as final.  The parameters of the best fits are  in Table 2.  The results of the  fits for the
$f(x)=\sech^{2}(x)$ case is presented in Figure 3. The corresponding
density profiles are shown in Figure 4.
There is not enough data to distinguish between the two functional
forms, but the resulting fit parameters are different in each case.
We prefer the $\sech^2(x)$ model because it does not have a midplane
cusp, and yields fewer outliers. For this case:

\begin{eqnarray}
n_e(r,z) &=& 1.77 \times 10^{-2} \cm^{-3}
          ~\frac{\sech^2(r/15.4 \kpc)}{\sech^2(R_\odot/15.4 \kpc)}
          ~\sech^2(z/1.10 \kpc) \nonumber \\
         &+& 1.07 \times 10^{-2} \cm^{-3}
          ~\frac{\sech^2(r/3.6 \kpc)}{\sech^2(R_\odot/3.6 \kpc)}
          ~\sech^2(z/0.04 \kpc) \nonumber
\end{eqnarray}

\noindent
These results are comparable to the previous axisymmetric model
of Cordes et al. (1991), although our thin disk component has a lower
midplane density ($n= 10^{-2}~{\rm cm^{-3}}$ vs. $n=20 \times
10^{-2}~{\rm cm^{-3}}$) and a shorter scaleheight ($h=40$ pc
vs. $h=175$ pc). We also find a noise parameter of $A=0.30$.  Savage
et al. (1990) did a similar study with a smaller sample of
pulsars. The value of the exponential scale height found is consistent
with theirs within the error bars, but their intrinsic scatter ($1.65
= 1+A$) is larger than ours.

\begin{figure}
\plotone{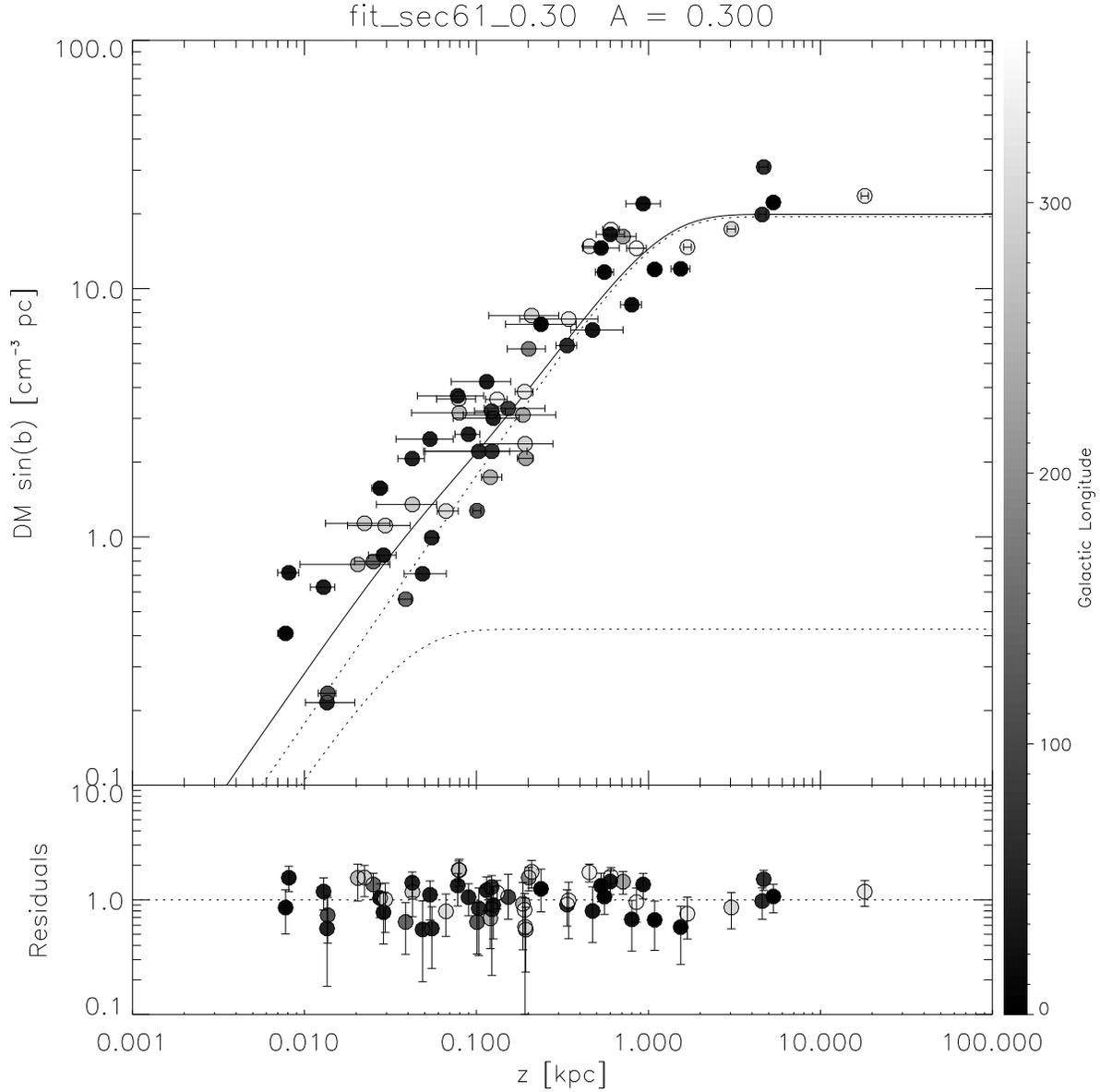}
\caption{[Upper panel] DMsin(b) as a function of
z for the pulsars used in fitting our final model. The solid line shows the two
component model with $f(x)=\sech^2(x)$ at the solar radius. The dotted
line shows the two individual components.  The error bars show only
the effects of distance uncertainty and do not incorporate the noise
parameter. [Lower panel] The residual values for our fit, defined as
$\DM_{data}/\DM_{model}$. The uncertainties incorporate the effects of
our noise parameter, $A=0.30$. No clear trend in the residuals with
galactic longitude (shown in grayscale) is observed. }
\end{figure}

\begin{figure}
\plotone{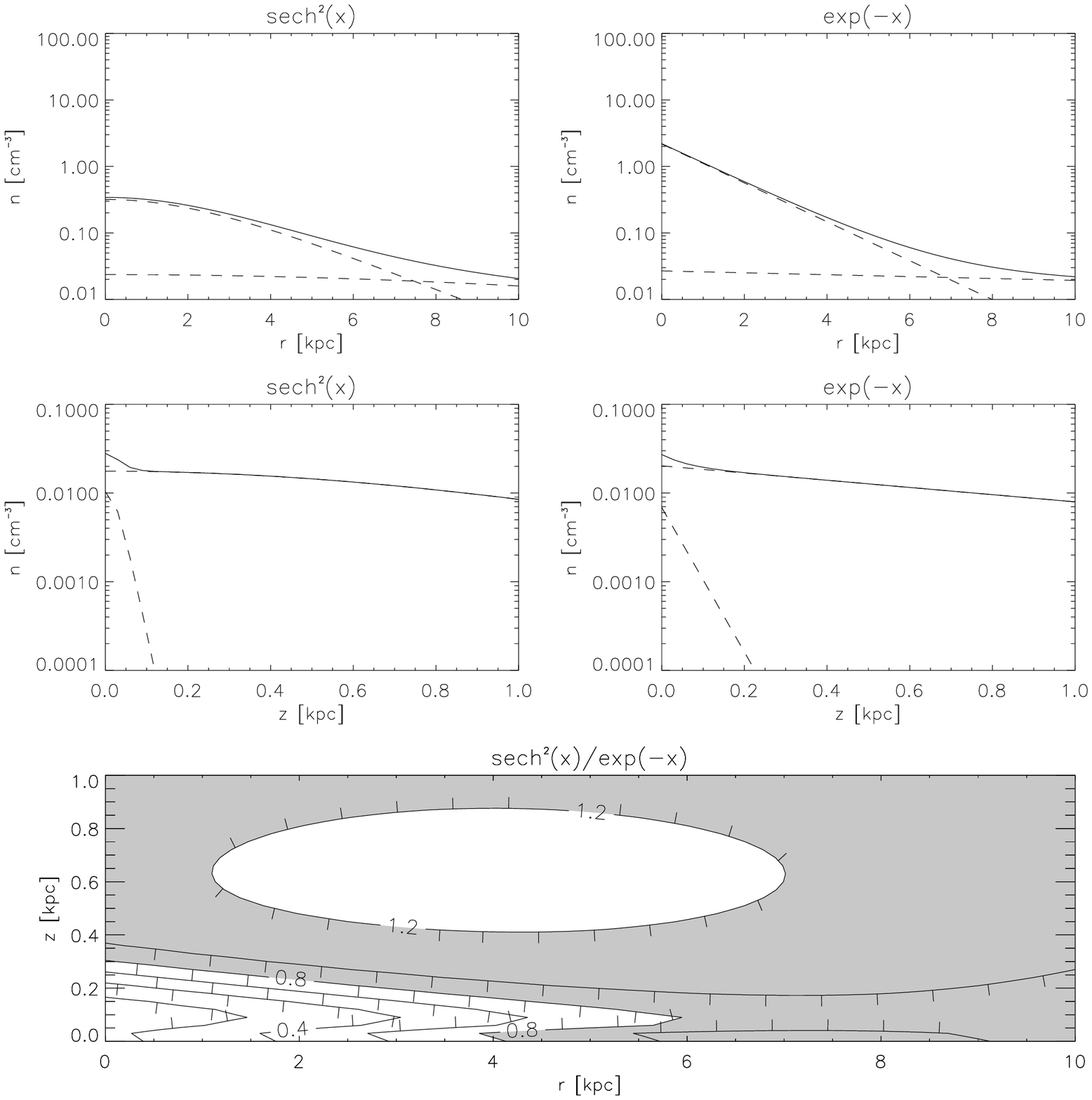}
\caption{Resulting density distributions, and
comparison of the two functional forms.  Dashed lines show the
individual components and the solid lines show the sum.  Upper panels
shows midplane density versus galactocentric radius. Middle panel
shows $n(z)$ for $r=r_\sun=8.5 \kpc$. The lower panel shows ratio of the
two model densities versus $r$ and $z$.  The pulsar dataset cannot
distinguish between the two functional forms. The shaded region shows
where the predicted electron density of the two models differ by less
than 20\%.}
\end{figure}

The procedure for spotting the outliers was the following: Consider
the values of $\DM_{model}$ obtained by integrating $n_e$ in the line
of sight towards each pulsar to the distance brackets, and call them
$\DM_+$ and $\DM_-$.  Consider now the values,

$$x_\pm = \frac{\DM_{data}}{\DM_{model}}
 \pm \sqrt{\left( \frac{\DM_{model} - \DM_{\pm}}{\DM_{model}} \right)^2
 +A^2}, $$

\noindent
for each pulsar (the error bars in the bottom panel of Figures 5 and 6
are the values of the square root above).  If $\sign(x_+ -1)=\sign(x_-
-1)$, then that pulsar is considered an outlier. As mentioned, the
pulsars spotted as outliers for both functional forms are taken out of
the sample for the final calculation of the fit.

\section{Deviations from the Smooth Axisymmetric Model}

Since observations of HII regions in the Galaxy show that there are
clearly inhomogeneities and asymmetries in the distribution of free
electrons, we have looked for patterns in the spatial and statistical
distribution of our residuals, $DM_{data}/DM_{model}$. We discuss in
turn, the individual outliers, the distribution of residuals with
respect to longitude and distance, and the nature of the scatter about
our smooth model. In the future, the combination of
this data with new radio recombination surveys for distant HII
regions and velocity-resolved $H\alpha$ surveys of more nearby gas
will yield a more complicated, but realistic, model.

\subsection{Outliers}

Of the 76 pulsars with both upper and lower limits, 15 are outliers in
both the exponential and $\sech^{2}(x)$ model. These outliers are noted
in Table 3, together with the observed dispersion measure, and the
dispersion measure that we would predict given the distance,
$DM_{\pm}=(1 \pm A)DM_{model}(D)$.   Two of these pulsars have dispersion
measure that are lower than one would expect given their distance. The
first, B1741-11, with a timing parallax distance, is only 0.36 kpc
distant. Given the lumpiness of the local interstellar medium (Cox \& Reynolds 1987; Toscano
et al. 1999), it is not out of the question for such a low density
sightline to arise for such a short distance. The second pulsar with a much lower dispersion
measure than expected is B1937+21. This pulsar, which has a kinematic
distance of $d=4.6$ kpc to 14.6 kpc, has $DM=71~ {\rm cm^{-3}~pc}$, while
our model yields $DM_{-}=208~ {\rm cm^{-3}~pc}$. This yields a mean of
electron density of $n_{e}< 0.016~{\rm cm^{-3}}$ over at least a 4.5 kpc
pathlength!  Further timing parallaxes for this pulsar could confirm
this unusual result.  

While low DM outliers are difficult to explain, high DM outliers are
likely to arise due to the passage the pulsar line of sight through a
dense HII region.  Of the 13 high outliers, four are associated with SNR
(Vela, MSH 15-52 SNR, G308.8-0.1, and W28) and one has a 10
$M_{\sun}$ companion (and presumably an associated HII region). Using the
ionizing output luminosity tabulated in Osterbrock (1989), the dispersion
measure of an HII region around an O9 star, for example, would be
$DM=2nR_{S}=315~ {\rm cm^{-3} pc}$, where $R_{S}$ is the Str\"{o}mgren
radius. The excess dispersion measure, defined as
$DM_{excess}=DM_{data}-DM_{+}$ for the 13 high outliers range from
$DM_{excess}=7 - 578~ {\rm cm^{-3} pc}$.  Thus these lines of sight are
consistent with the intersection of the line of sight with discrete HII
regions.  However, we have searched catalogs of diffuse HII regions
(Lockman, Pisano, and Howard 1996) and the WHAM maps (Haffner 2000) for
correlations with these northern declination pulsars in this sample, but
nothing outstanding was found. Since the majority of these pulsars lie at
southern declinations, the high angular resolution $H\alpha$ maps of
Gaustad et al. (1997) will be extremely useful in the future.  

There are two outliers for which we suspect the distance estimate may
be incorrect. The distance to B0656+14 was obtained using an X-ray
luminosity distance. Given the number of assumptions necessary to
estimate the X-ray luminosity of a neutron star, we have some concerns
about the reliability of this method.  The distance to B0823+26 is
based on a parallax measurement by Gwinn et al. (1986). Since the other
pulsar examined in this study (B0950+08) has had a significant
revision in its distance, a reconsideration of this pulsar parallax
may be in order.

We have also compared to the 29 pulsars for which there are only upper
or lower limits. We found that 26 of the limits are satisfied by the
model while B2020+28 ($D> 3.1$ kpc), B2016+28 ($D>3.2$ kpc), and
B1818-04 ($D<1.6$ kpc) are not.  Thus our model satisfies the distance
constraints of 91 out of 109 pulsars.

\subsection{Spatial Distribution of Residuals}

We now consider whether the known asymmetries in the distribution of
galactic HII regions is reflected in the current dataset. A plot of
$\DM_{data}/\DM_{model}$ versus galactic location is presented in
Figure 5, with the spiral arm positions used by TC overlaid. There
seem to be two lines of pulsars with a higher than expected dispersion
measure, marked by dashed lines. Some of these pulsars have been
discussed by Johnston et al. (2001) as particularly noticeable
outliers. One of these groups agrees roughly with the position of one of the
spiral arms and has a pitch angle $27^{\circ}$ from the tangent. The
other has a pitch angle of $22^{\circ}$ and is not coincident with
any of the spiral arms.  Given the distance uncertainties for these
pulsars, it seems clear that any spiral structure that might exist is
only weakly exhibited in this data set.

\begin{figure}
\plotone{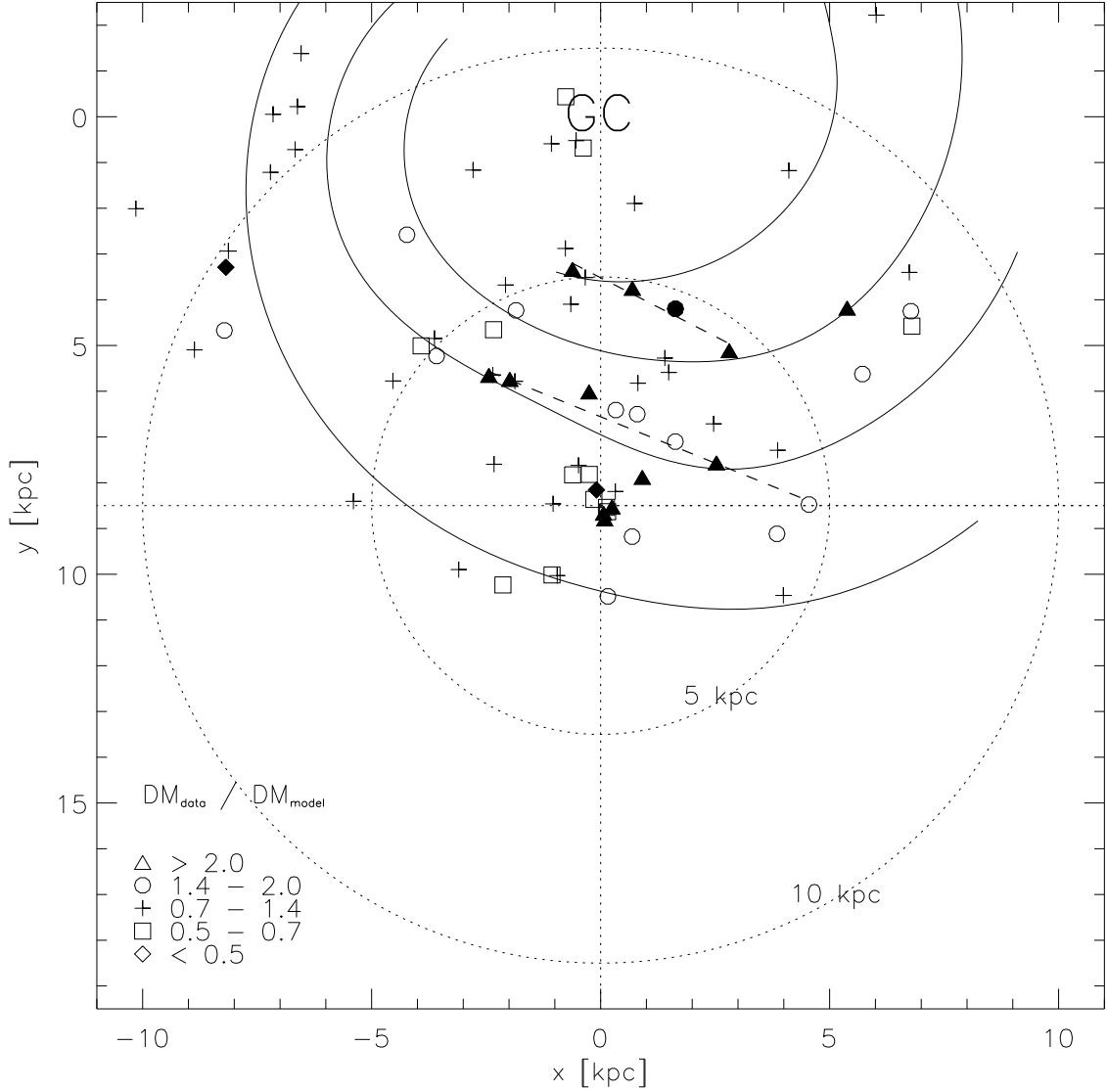}
\caption{$\DM_{data}/\DM_{model}$ ratio (values shown via legend symbol)
for pulsars with $|z| < 300 \pc$ projected on the Galactic plane.
The grid circles are labeled with distance from the Sun. The solid
lines trace the center of the spiral arms in the TC model.
Pulsars which are identified 
as outliers in Table 3 have filled symbols. There are two regions 
of high $\DM_{data}/\DM_{model}$, at approximately 2 and $5 \kpc$
from the Sun towards the galactic center.
One of these regions (noted with dashed lines)
coincides with the position of the Sagittarius-Carina arm.}
\end{figure}

We have also considered whether there is evidence for a difference in
the estimated midplane density if we use only pulsars identified as
``interarm pulsars'' to estimate the midplane density at the solar
neighborhood, where those pulsars are marked as such in Figure 1. We
found that there was a slight decrease in the derived midplane density,
a factor of $2/3$, compared to the total dataset. However, some of these
pulsars may lie in or beyond the Local Arm, which although not included
in the TC model, is known to exist in the $H\alpha$ data (Reynolds 1983). 

\subsection{Constraints on Clumpiness}

We now consider what factors affect the scatter in the relationship
between our simple axisymmetric model and the observed data. Figure 6
shows the comparison between the $\DM_{model}$ and
$\DM_{data}$. $\DM_{model}$ takes into account the geometry of the
distribution, so it measures the effective integration path. Therefore,
we will use it instead of the distance in order to examine the nature of
the scatter. An interesting feature in this plot is that the scatter
appears to be a fixed fraction of the total dispersion measure.

\begin{figure}
\plotone{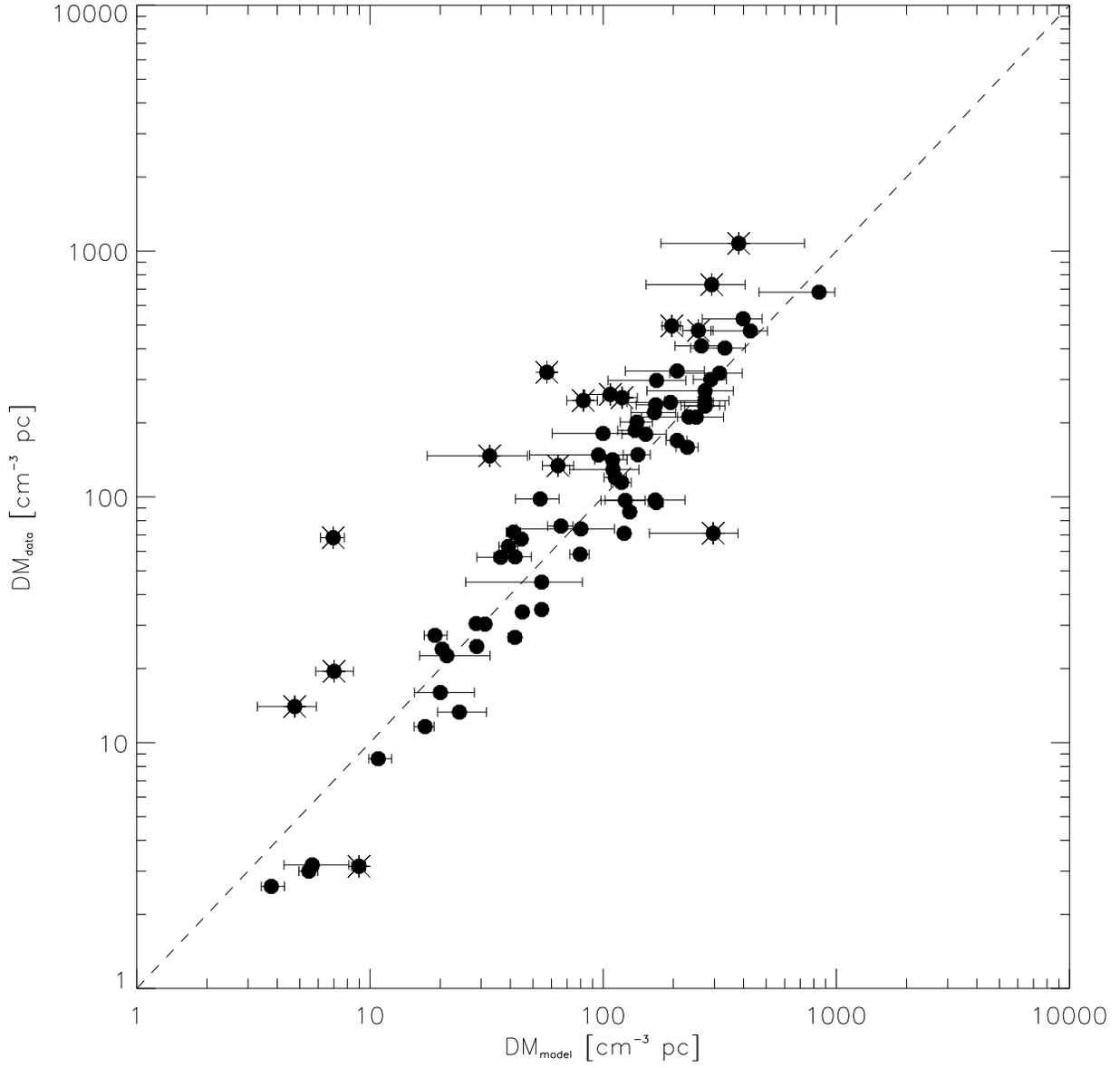}
\caption{$\DM_{data}$ vs. $\DM_{model}$ for the best fit model with
$f(x)=\sech^{2}(x)$. The starred points are the pulsars identified as
outliers. The error bars shown here only show the uncertainties in the
distances and do not incorporate our parameterization of scatter in
the relationship.}
\end{figure}

We considered the possibility that the scatter of $\DM_{data}$
about the smooth model might derive from a patchiness of the
distribution of electrons in the Galaxy. Such a patchiness of the diffuse
ionized medium has been predicted, for example, by Miller \& Cox (1993)
using the observed locations of O stars in the Solar Neighborhood, and
a model for the ISM distribution, to calculate the steady state Str\"omgren
volume distribution and ionization.

What would happen if the warm ionized medium were {\it purely} located in
discrete lumps (or HII regions)?\footnote{Although some authors argue
that there is a continous of power in all scales, here we are considering
patches of ionization of finite size.}
In that case, we define
$\widehat{DM}=DM_{model}$ as the dispersion measure that would result for
the average line of sight through some variable number of clumps. The
expected total number of lumps intersected along a line of sight would
then be $n=\widehat{DM}/DM_{lump}$. The variance in observed number should
also be $n$ so that 
$(DM_{data}-\widehat{DM})^{2}=nDM_{lump}^{2}=\widehat{DM} \cdot DM_{lump}$.
We can therefore define a quantity, $\mu$, for each pulsar, 

\begin{equation}
\mu = \frac{(\DM_{data}-\widehat{DM})^2}{\widehat{DM}}~.
\end{equation}

\noindent
If the lump sizes are normally distributed, $\mu$ should be
independent of $\widehat{DM}$, and its average over a large enough
sample of pulsars should be the dispersion measure of the lump. In
Figure 7 we plot the running mean of this quantity for both the
top-down (from large $DM_{model}$ to small $DM_{model}$) and bottom-up
sums. There is a strong trend in the lump size estimator with the distance.
This could be explained by having two lump populations: small frequent
lumps and large rarer lumps.  Nearby, we pick only small lumps,
yielding a small mean.  As we move farther, we pick up more large
lumps and the mean value increases.  This could explain the steps
observed in the bottom-up running mean, while the top-down running
mean is flatter. We thought that a log-normal distribution with the
appropriate shape parameter might have that property, but found that
 a log-normal distribution could almost reproduce the properties of
Figure 5 (constant fractional scatter with increasing DM), but not
Figure 7 ($\mu$ is not constant).

\begin{figure}
\plotone{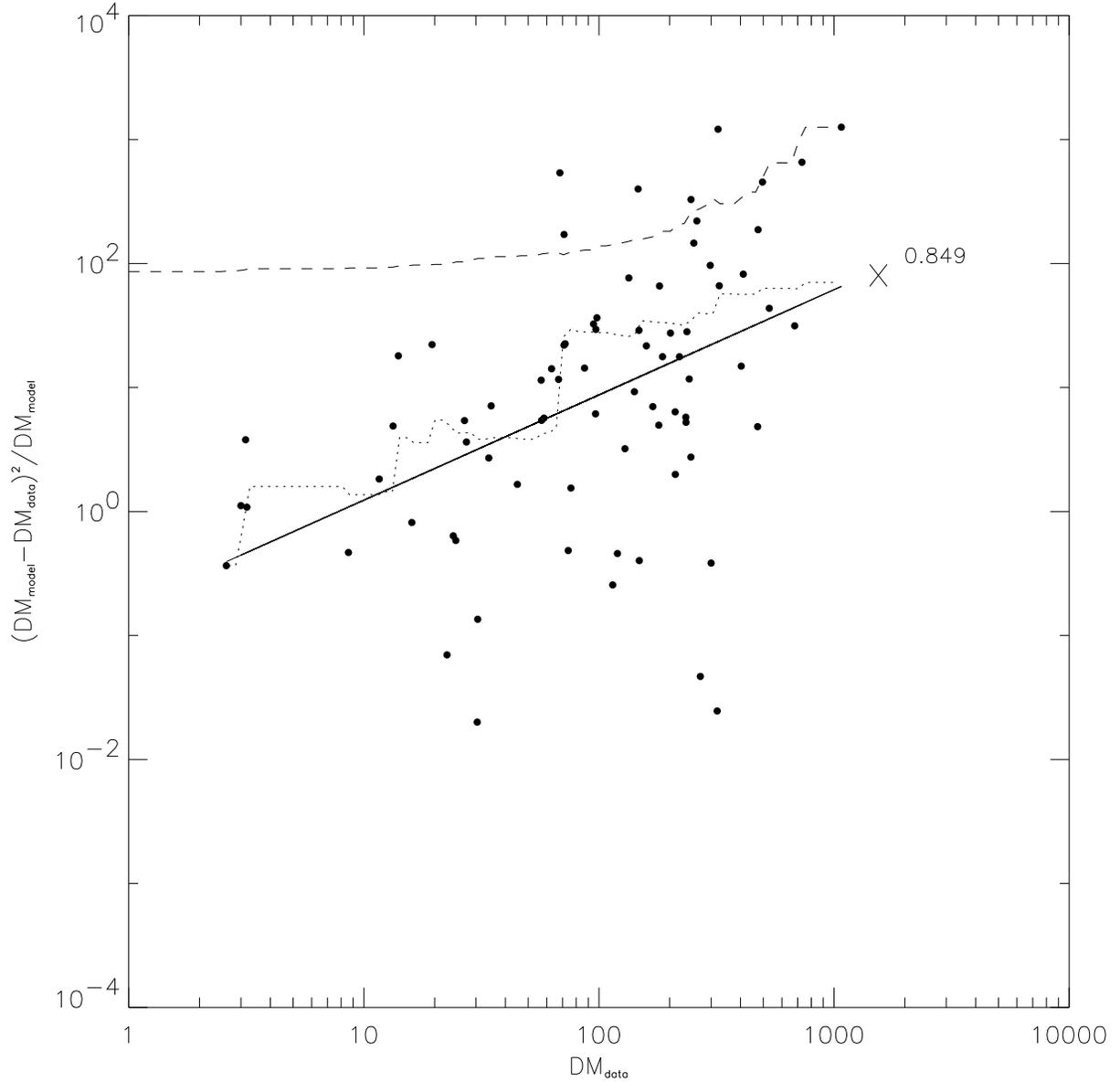}
\caption{Estimation of the mean lump size. The top-down (dashed) and
bottom-up (dotted) running averages are shown.  If the variance where
due to random encounter of lumps along the line of sight, the average
of the ordinate should be constant, independent of DM, and roughly
equal to the lump size.  The solid line is the least-squares fit to
the logarithm of the data. }
\end{figure}


Another possibility is that the deviations of the observed and modeled DM
are not due to statistical noise, but instead fractional errors in the
distance measurements. In such case, the variance is proportional to
$f^{2}\widehat{DM}^2$, where $f$ is the approximate fractional error,
rather than $\sqrt{n}DM_{lump}$. In this case, a plot of
$(\widehat{DM}-\DM_{data})^{2}/\widehat{DM}^{2}$, should be roughly flat,
which is verified in Figure 8.  The  corresponding value is $f \cong
30\%$.  If distance uncertainties are indeed the main source of scatter,
it will be difficult  to say anything definitive about the lumpiness of
the warm ionized medium based on this type of data.

\begin{figure}
\plotone{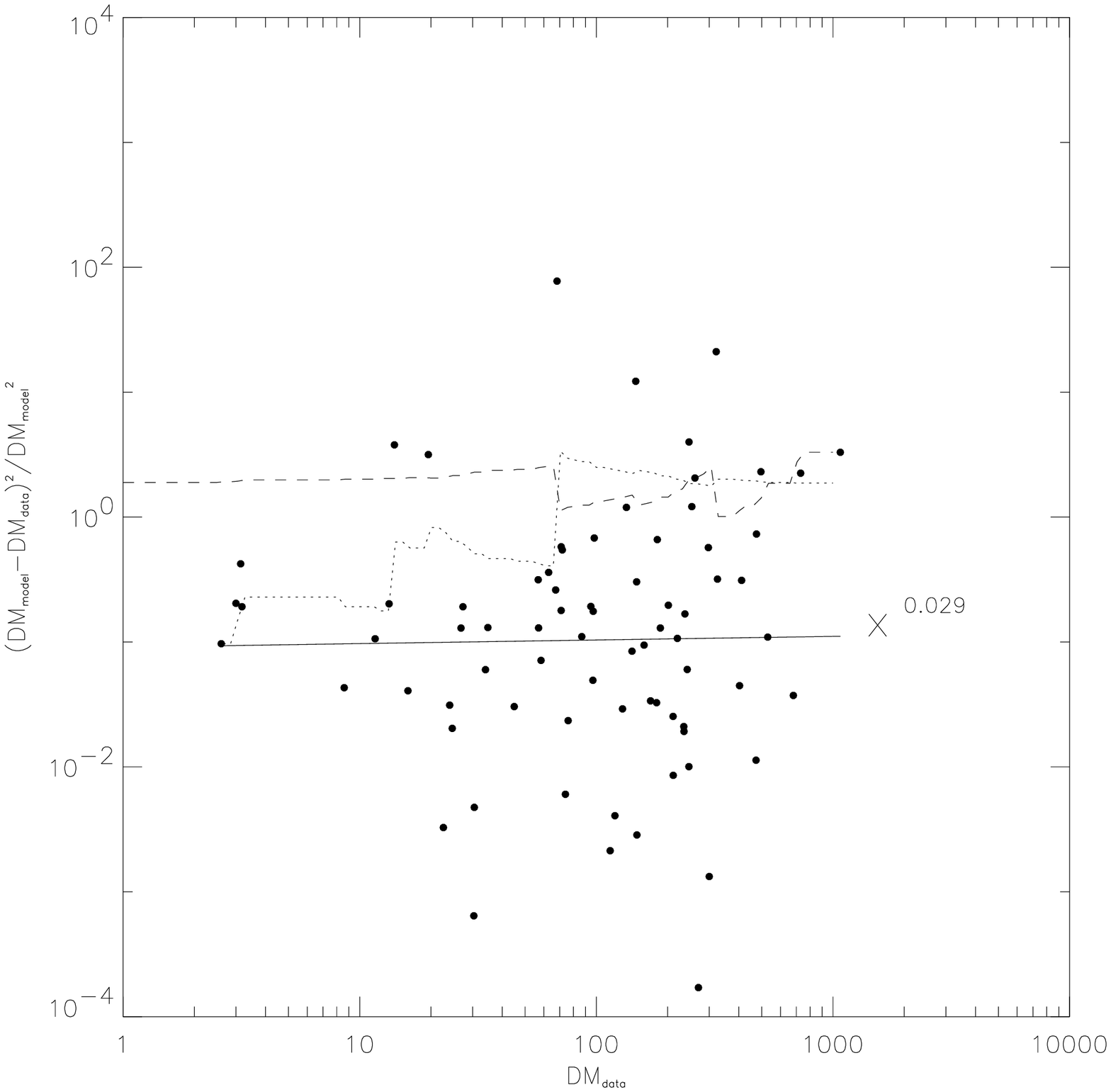}
\caption{Estimation of the square of the mean
fractional error. The top-down (dashed) and bottom-up (dotted) running
averages are shown. The solid line is the least-squares fit to logarithm of the
data.  Since the slope is close to zero, this implies that the
dispersion is approximately a constant fraction of DM. We suspect that
this behavior is due principally to distance uncertainties.}
\end{figure}

\section{Predicting Pulsar Distances}

One of the principal uses for a model of the galactic free electron
distribution model is to predict the distance to pulsars. While we
have not yet introduced the effects of asymmetries, spiral structure,
and individual H II regions, we have written two FORTRAN routines (one
for each functional form tested) that calculate pulsar distances using
the model parameters in Table 2.\footnote{These programs may be
obtained by contacting the authors or at the web site {\tt
http://wisp5.physics.wisc.edu/$\sim$gomez/publica.html}.}
A comparison of the model distances
and true distances for our sample of pulsars is given in Figure 9,
using the $f(x)=\sech^2(x)$ model.  The error bars are obtained by
calculating the distance that corresponds to $(1 \pm A)DM_{data}$. We
note that no pulsars have a DM higher than the asymptotic limit when
the uncertainty associated with our noise parameter $A$ is considered.
 
\begin{figure}
\plotone{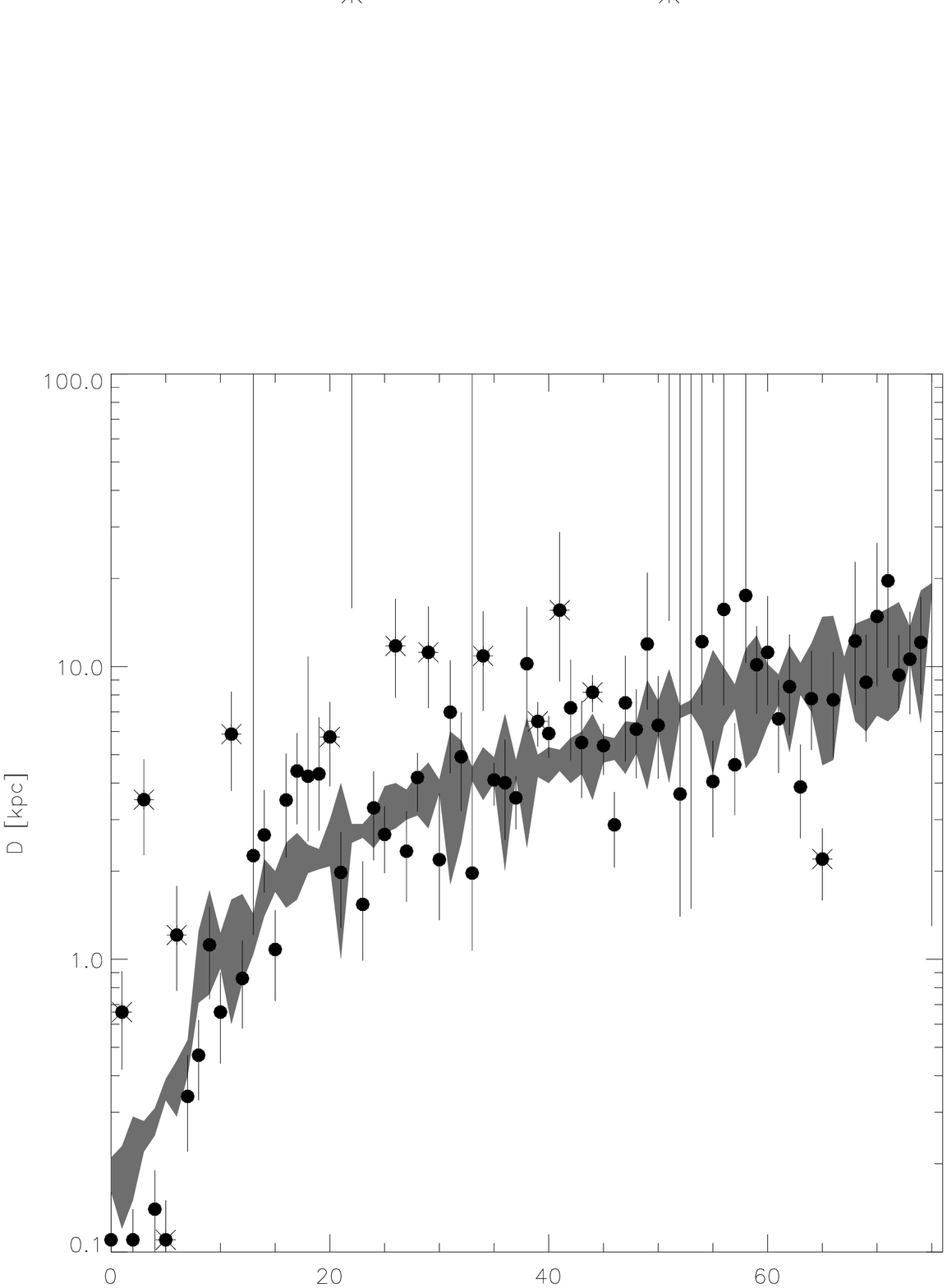}
\caption{Comparison of the predicted and
measured distances using the $f(x)=\sech^{2}(x)$ model.  The
horizontal axis is the pulsar number, sorted by distance.  The gray
region is the quoted range in measured distances.  The starred points
are the outlier pulsars identified in Table 3, and the error bars include
the effect of the noise parameter, $A=0.3$.}
\end{figure}

In Figure 10, we compare the distances predicted by the TC model and
our model with the observed distance constraints. When we consider
only those pulsars with allowed DM (smaller than the asymptotic
value), the dispersion in our model is similar to the model of TC, but
with fewer free parameters. 

\begin{figure}
\plotone{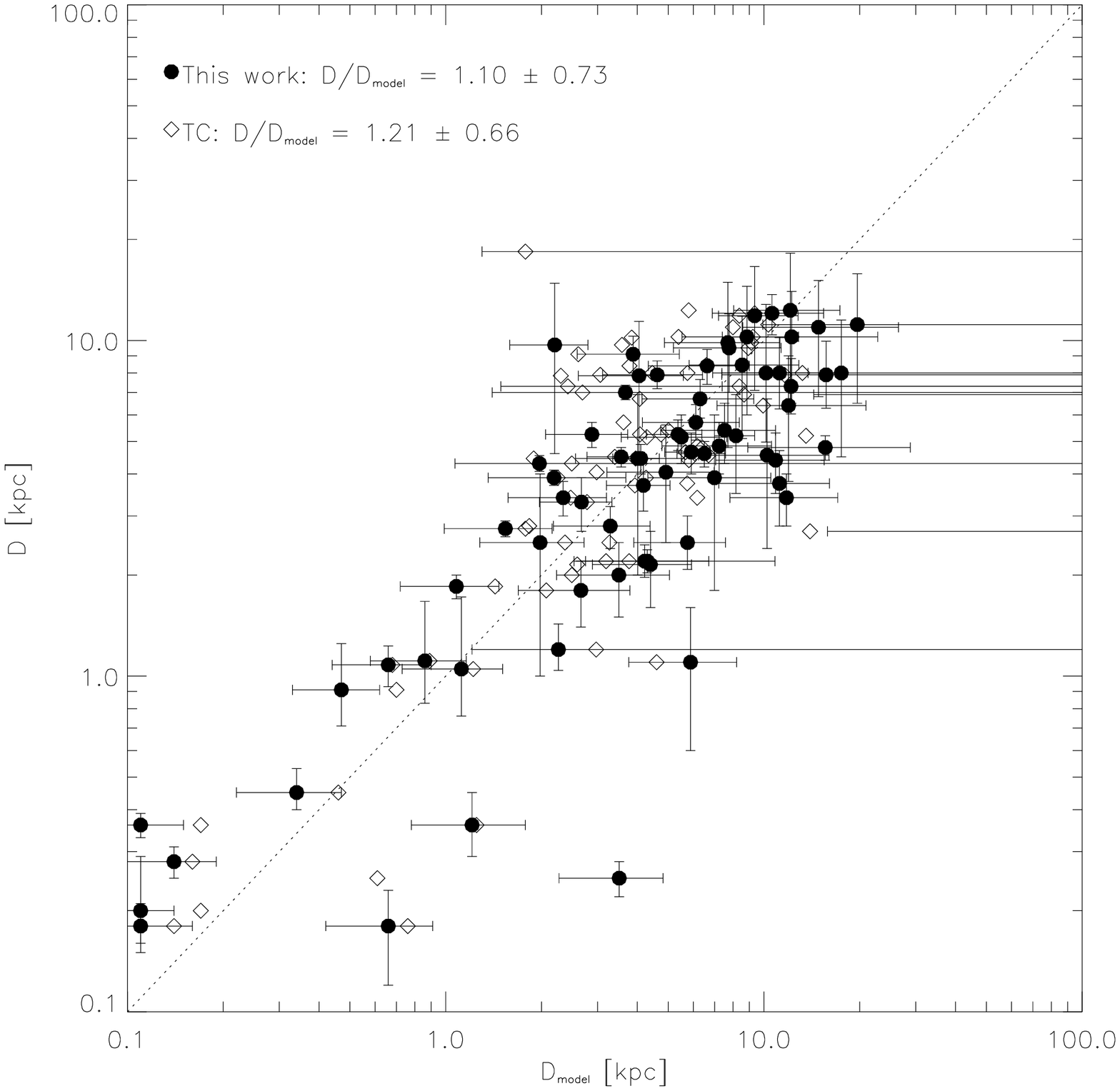}
\caption{Comparison of the predicted and
measured distances using both our $f(x)=\sech^{2}(x)$ model and the
Taylor-Cordes model.  The horizontal error bars are computed by
estimating the distance corresponding to values of $\DM = (1 \pm A)
\DM_{data}$. The dispersion in our model is similar to the TC
model, despite the fact that we have nine fewer adjustable parameters.}
\end{figure}


We note however that the model we have developed is relatively
unconstrained for pulsars interior to a galactocentric radius
of $R \sim 4$ kpc and exterior to 12 kpc. For example, unlike Taylor
\& Cordes, we have not included a annulus of electron density at
$R=4$ kpc, which presumably would be associated with the molecular
ring. Lazio
\& Cordes (1998a,c) have discussed how additional information can be
used to constrain these regions. We intend to address these
issues in the future when we address the non-axisymmetric structure using
the Wisconsin H$\alpha$ Survey. 

\section{Conclusions.}

A smooth model for the distribution of galactic free electrons was
obtained from a set of 109 pulsars with independent distance
information.  Although a more complex model incorporating spiral arms
might be possible, we do not think that it is well constrained by this
pulsar data alone, so we chose to use a simpler and probably more robust
functional form.  The exponential scale height obtained is consistent
with the value quoted in Reynolds (1996). The scatter parameter found
($A=0.3$) is smaller than the one found by Savage et al. (1990).  This
scatter parameter is used to predict a range of confidence in the
predicted distances to pulsars.

Of pulsars with both upper and lower distance limits, fifteen pulsars are
identified as outliers, with thirteen of these showing excess dispersion
measure. Some of these are associated with supernova remnants or known
HII regions. There is one very unusual pulsar, $B1937+21$, with an
extremely low dispersion measure given its distance. In examining the
residuals, we identified two regions of enhanced electron density, one of
which corresponds well with the expected position of the
Sagittarius-Carina spiral arm.  

We found that a simple probabilistic model for a lumpy WIM failed to
reproduce the deviations of the observed data from the smooth model.  We
suspect that the main source of scatter in our model is due to distance
uncertainties, although it seems clear that are also occasionally large
anomalous dispersion measures associated with HII regions. Some of these
are in spiral arms, but their distribution may not be uniform in these
arms. 

\acknowledgments We are grateful to Ron Reynolds, Matt Haffner, Joel Weisberg, and Linda Sparke for useful advice and encouragement, to NASA ATP
grant NAG5-8417 for financial support, and to CONACYT-MEXICO for support for GCG.

\makeatletter
\def\jnl@aj{AJ}
\ifx\revtex@jnl\jnl@aj\let\tablebreak=\\\fi
\makeatother


\begin{deluxetable}{rrrrrrrrll}
\tabletypesize{\scriptsize}
\tablecolumns{10}
\tablecaption{Pulsars with independent distance information}
\tablehead{\colhead{PSR} &  
\colhead{l}      & 
\colhead{b}      & 
\colhead{DM}  &
\colhead{$D_{min}$} &
\colhead{D} &
\colhead{$D_{max}$} &
\colhead{$\frac{<n_{e}>}{10^{-3}}$} &
\colhead{Method\tablenotemark{1}} &
\colhead{Refs} 
}
\startdata
 $0435-47  \tablenotemark{*}    $  & 253.40 &	-42.00  & 2.60      &	0.16    &	0.18    & 0.21 &   14.4   &    PDD               &  46   \\
 $0656+14  \tablenotemark{*}  $  & 201.11 &	8.26    & 14.02     & 0.12    &	0.18    & 0.23 &   77.9   &    X               &  49   \\     
 $1929+10\tablenotemark{\dag} $  & 47.38  &	-3.88   & 3.18      & 0.15    &	0.20    & 0.29 &   15.9   & ${\bf \Pi}$,K   &5,8,17,27,28,{\bf 39}\\     
 $0833-45\tablenotemark{\dag} $  & 263.55 &	-2.79   & 68.20     & 0.22    &	0.25    & 0.28 &  272.8   &    {\bf Vela SNR},$\Pi$        & 47,{\bf 58}   \\    
 $0950+08\tablenotemark{\dag} $  & 228.90 &	43.70   & 3.00      &0.25    & 0.28    & 0.31 &   10.7     &    $\Pi$           & 30,{\bf 52} \\
 $1741-11  \tablenotemark{*}    $  &  14.79 &    9.18   & 3.14      & 0.33      & 0.36    & 0.39 &    8.7   &    TP               & 51   \\
 $0823+26                  $  & 197.00 &	31.70   & 19.50     &	0.29    &	0.36    & 0.45 &   54.2   &    $\Pi$           & 30  \\     
 $1451-68                       $  & 313.90 &	-8.50   & 8.60      &	0.40    &	0.45    & 0.53 &   19.1   &    $\Pi$           & 33  \\     
 $1855+09 \tablenotemark{\dag}$  & 42.29  &	3.06    & 13.31     &0.71    &	0.91    & 1.25 &   14.6   &    {\bf TP}, K      & 23, 34, {\bf38} \\   
 $2021+51  \tablenotemark{*}    $  & 87.86  &	8.38    & 22.58     &	0.76    &	1.05    & 1.72 &   21.5   &    $\Pi$           & 42   \\     
 $1534+12  \tablenotemark{*}    $  & 20.00  &	47.80   & 11.62     &0.93    &	1.08    & 1.23 &   10.8   &    PDD,TP           & 50    \\     
 $1259-63  \tablenotemark{*}    $  & 304.2  &  -0.992   & 146.72    &   0.60    &       1.10    & 1.60 &  133.4   &    SP              & 59 \\
 $1711+07  \tablenotemark{*}    $  & 28.75  &	25.22   & 15.99     &	0.83    &	1.11    & 1.67 &   14.4   &    TP               & 37   \\     
 $0919+06  \tablenotemark{*}  $  & 225.42 &	36.39   & 27.31     &1.04    &	1.20    & 1.43 &    22.7   &    $\Pi$           & {\bf 53},48\\     
 $0355+54                       $  & 148.20 &	0.80    & 57.00     &	1.40    &	1.80    & 2.20 &   31.7   &    K               & 11,12,15  \\     
 $0329+54                       $  & 145.00 &	-1.20   & 26.80     &	1.70    &	1.85    & 2.00 &   14.5   &    K               & 1,2,3,9,15,24\\     
 $0531+21                       $  & 184.56 &	-5.78   & 56.79     &	1.50    &	2.00    & 2.50 &   28.4   &    Crab SNR        & 25  \\     
 $1358-63  \tablenotemark{*}    $  & 310.60 &	-2.10   & 98.00     &	1.60    &	2.15    & 2.70 &   45.6   &    K               & 43  \\     
 $1620-26  \tablenotemark{\dag} $  & 350.98 &	15.96   & 62.86     &	1.97    &	2.20    & 2.46 &   28.6   &    NGC 6121 (M4)   & 56,60  \\     
 $1740-53  \tablenotemark{*}    $  & 338.20 &-11.90   & 71.80     & 2.12    & 2.30    & 2.49 &    31.2   &    NGC 6397        & 56,55,61  \\
 $1951+32                       $  & 68.77  &	2.82    & 44.98     &	1.00    &	2.50    & 4.00 &   18.0   &    CTB80 SNR       & 29    \\   
 $1807-24  \tablenotemark{*}    $  &   5.80 & -2.20   & 134.00    & 2.17    & 2.60    & 3.11 &    51.5  &    NGC 6544        & 56,55,62  \\ 
 $1054-62 \tablenotemark{\dag}$  & 290.30 &	-3.00   & 321.00    &2.50    &	2.70    & 2.90 &  118.9   &    K               & 18,22,{\bf40}\\     
 $0138+59                       $  & 129.10 &	-2.10   & 34.80     &	2.60    &	2.75    & 2.90 &   12.7   &    K               & 12 \\     
 $1706-44  \tablenotemark{*}    $  & 343.10 &	-2.70   & 76.00     &	2.40    &	2.80    & 3.20 &   27.1   &    K               & 40 \\     
 $1853+01                       $  & 34.56  &	-0.50   & 96.70     &	2.70    &	3.30    & 3.90 &   29.3   &    W44 SNR         & 26 \\     
 $1900+01                       $  & 35.70  &	-2.00   & 246.40    &	2.80    &	3.40    & 4.00 &   72.5   &    K               & 16 \\     
 $2334+61  \tablenotemark{*}    $  & 114.28 &	0.23    & 58.38     &	3.00    &	3.40    & 3.80 &   17.2   &    G114.3+0.3 SNR  & 45  \\     
 $1900+05                       $  & 39.50  &	0.20    & 179.70    &	3.10    &	3.70    & 4.30 &   48.6   &    K               & 24 \\     
 $1859+07                       $  & 40.60  &	1.10    & 261.00    &	2.80    &	3.75    & 4.70 &   69.6   &    K               & 21 \\     
 $0835-41 \tablenotemark{\dag}$  & 260.90 &	-0.30   & 148.00    &1.80    &	3.90    & 6.00 &   37.9   &    K               & 13,{\bf 43}\\  
$1910-59 \tablenotemark{*}     $  & 336.5  & -25.60  &  34.00    & 3.79    & 4.00    & 4.22 &    8.5   &    NGC 6752        & 56,55,63  \\   
 $1046-58  \tablenotemark{*}    $  & 287.40 &	0.60    & 129.00    &	2.50    &	4.05    & 5.60 &   31.9   &    K               &  43   \\     
    
 $1509-58                       $  & 320.32 &	-1.16   & 253.20    &	3.50    &	4.40    & 5.30 &   57.5   &    MSH15-52 SNR    & 26    \\     
 $1800-21                       $  & 8.40   &	0.10    & 234.20    &	4.00    &	4.45    & 4.90 &   52.6   &    K,G8.7-0.1 SNR              & 24,57   \\     
 $0740-28\tablenotemark{\dag} $  & 243.80 &	-2.40   & 74.00     &2.00    &	4.45    & 6.90 &   16.6   &    K               & 9,10,{\bf 40}\\   
$0021-72C \tablenotemark{\dag} $  & 305.92 &	-44.89  & 24.61     &	4.27    &	4.50    & 4.75 &    5.5   &    NGC 104 (47 Tuc)& 56,64   \\    
 $1845-01                       $  & 31.30  &	0.00    & 159.10    &	4.20    &	4.50    & 4.80 &   35.3   &    K               & 18,20 \\     
 $0906-49  \tablenotemark{*}    $  & 270.30 &	-1.00   & 181.00    &	2.40    &	4.55    & 6.70 &   39.8   &    K               & 40   \\     
 $1641-45                       $  & 339.20 &	-0.20   & 475.00    &	4.20    &	4.60    & 5.00 &  103.3   &    K               & 14,24  \\     
 $1830-08  \tablenotemark{*}    $  & 23.40  &	0.10    & 411.00    &	4.00    &	4.65    & 5.30 &   88.4   &    K               & 41   \\    
 $1718-35  \tablenotemark{*}    $  & 351.70 &	0.70    & 496.00    &	4.40    & 4.80    & 5.20 &  103.3   &    K               & 41   \\     
 $1914+13                       $  & 47.60  &	0.50    & 236.80    &	4.00    &	4.85    & 5.70 &   48.8   &    K               & 20   \\     
 $1907+10                       $  & 44.80  &	1.00    & 148.40    &	4.30    &	5.15    & 6.00 &   28.8   &    K               & 20  \\     
 $1758-23  \tablenotemark{*}    $  & 6.80   &	-0.10   & 1074.00   &	3.50    &	5.20    & 6.90 &  206.5   &    K, W28 SNR               & 36    \\     
 $1829-08                       $  & 23.30  &	0.30    & 300.00    &	4.70    &	5.25    & 5.80 &   57.1   &    K               & 24  \\     
 $1915+13                       $  & 48.30  &	0.60    & 94.80     &	4.80    &	5.25    & 5.70 &   18.1   &    K               & 20  \\      
 $2111+46                       $  & 89.00  &	-1.30   & 141.50    &	4.30    &	5.40    & 6.50 &   26.2   &    K               & 12  \\     
 $1821-24  \tablenotemark{\dag} $  & 7.80   &	-5.58   & 119.83    &	5.03    &	5.70    & 6.46 &   21.0   &    NGC 6626 (M28)  & 56,65   \\     
 $1154-62  \tablenotemark{*}    $  & 296.70 &	-0.20   & 325.00    &	3.80    &	6.40    & 9.00 &   50.8   &    K               & 43   \\     
 $1701-30  \tablenotemark{*}    $  & 353.60 &  7.30   & 114.40    & 6.04    & 6.90    & 7.88 &    16.6   &    NGC 6266 (M62)  & 56,55,66   \\
 $1338-62  \tablenotemark{*}    $  & 308.73 &	-0.04   & 730.00    &	4.00    &	6.90    & 9.80 &  105.8   &    G308.8-0.1 SNR  & 35    \\     
 $1908+00  \tablenotemark{\dag} $  & 35.54  &	-4.71   & 201.50    &	6.13    &	7.40    & 8.93 &   27.2   &    NGC 6760        & 56,67    \\     
 $1516+02B \tablenotemark{\dag} $  & 3.86   &	46.80   & 30.50     &	7.12    &	7.50    & 7.90 &    4.1   &    NGC 5904 (M5)   & 56,68,69     \\  
$1744-24A \tablenotemark{\dag} $  & 3.84   &	1.70    & 242.14    &	4.69    &	7.60    & 12.31 &   31.9   &    Ter 5           & 56,70   \\    
 $1639+36A \tablenotemark{\dag} $  & 59.00  &	40.91   & 30.36     &	7.33    &	7.70    & 8.09 &    3.9   &    NGC 6205 (M13)  & 56,71    \\      
 $1221-63  \tablenotemark{*}    $  & 300.00 &	-1.40   & 97.00     &	4.30    &	7.85    & 11.4 &   12.4   &    K               & 43   \\ 
 $1820-30A \tablenotemark{\dag} $  & 2.79   &	-7.91   & 86.80     &	7.26    &	8.00    & 8.82 &   10.9   &    NGC 6624        & 56,72   \\     
 $1240-64                       $  & 302.10 &	-1.50   & 297.40    &	4.50    &	8.00    & 11.5 &   37.2   &    K               & 14,22 \\      
 $1802-07  \tablenotemark{\dag} $  & 20.79  &	6.77    & 186.38    &	6.71    &	8.40    & 10.52 &   22.2   &    NGC 6539        & 56,73   \\        
 $1745-20  \tablenotemark{\dag} $  & 7.73   &	3.80    & 220.00    &	6.59    &	8.40    & 10.71 &   26.2   &    NGC 6440        & 56,74   \\   
 $1558-50 \tablenotemark{\dag}  $  & 330.70 &	1.30    & 169.50    &	7.40    &	8.40    & 9.40 &   20.2   &    K               & 20, 54 \\        
 $1323-62                       $  & 307.10 &	0.20    & 318.40    &	5.10    &	8.45    & 11.8 &   37.7   &    K               & 14  \\      
 $1718-19  \tablenotemark{\dag} $  & 4.87   &	9.74    & 71.00     &	7.55    &	8.60    & 9.80 &    8.3   &    NGC 6342        & 56,67    \\      
 $2002+31                       $  & 69.00  &	0.00    & 234.70    &	7.00    &	9.50    & 12.0 &   24.7   &    K               & 16  \\   
 $1937+21                     $  & 57.51  &	-0.29   & 71.04     &4.60    &	9.70    & 14.8 &    7.3   &  {\bf K},TP         & 19,38 \\        
 $1929+20                       $  & 55.60  &	0.60    & 211.00    &	4.80    &	9.85    & 14.9 &   21.4   &    K               & 24  \\      
 $1904+06                       $  & 40.60  &	-0.30   & 473.00    &	6.50    &	10.25   & 14.0 &   46.1   &    K               & 21  \\      
 $1913+10                       $  & 44.70  &	-0.70   & 246.10    &	6.00    &	10.25   & 14.5 &   24.0   &    K               & 24  \\ 
 $2127+11A \tablenotemark{\dag} $  & 65.01  &	-27.31  & 67.31     &	9.66    &	10.30   & 10.99 &   6.5   &    NGC 7078        & 56,75    \\      
 $1859+03                       $  & 37.20  &	-0.60   & 402.90    &	6.80    &	10.95   & 15.1 &   36.8   &    K               & 14,16\\      
 $1900+06                       $  & 39.90  &	0.40    & 530.00    &	6.50    &	11.15   & 15.8 &   47.5   &    K               & 24  \\      
 $1849+00                       $  & 33.50  &	0.00    & 680.00    &	7.10    &	11.85   & 16.6 &   57.4   &    K               & 21  \\      
 $1930+22                       $  & 57.40  &	1.60    & 211.30    &	10.40   &	12.05   & 13.7 &   17.5   &    K               & 24  \\ 
 $1557-50 \tablenotemark{\dag}  $  & 330.70 &	1.60    & 270.00    &	6.40    &	12.30    & 18.2 &   22.0   &    K               &14,18,54\\      
 $1310+18  \tablenotemark{\dag} $  & 332.96 &	79.77   & 24.00     &	17.41  &	18.30   & 19.23 &    1.3   &    NGC 5024 (M53)  & 56,76     \\      
 $0456-69                       $  & 281.20 &	-35.19  & 91.00     &	46.00   &	49.40   & 52.8 &    1.8   &    LMC             & 32    \\      
 $0502-66                       $  & 277.03 &	-35.50  & 65.00     &	46.00   &	49.40   & 52.8 &    1.3   &    LMC?            & 32   \\      
 $0529-66                       $  & 277.02 &	-32.80  & 100.00    &	46.00   &	49.40   & 52.8 &    2.0   &    LMC             & 32    \\      
 $0042-73                       $  & 303.51 &	-43.80  & 105.40    &	52.80   &	57.00   & 61.2 &    1.8   &    SMC             & 32    \\      
 $1749-28                       $  & 1.50   &	-1.00   & 50.90     &	0.13    &	        &      & $<391.5$ &  K                 & 2,3  \\      
 $1857-26  \tablenotemark{*}    $  & 10.34  &	-13.45  & 38.06     &	0.91    &	        &      & $< 41.8$ &  $\Pi$             & 48   \\      
 $1804-08                       $  & 20.10  &	5.60    & 112.80    &	1.50    &	        &      & $< 75.2$ &  K                 & 24  \\      
 $1821+05                       $  & 35.00  &	8.90    & 67.50     &	1.60    &	        &      & $< 42.2$ &  K                 & 20,21,24 \\      
 $1920+21                       $  & 55.30  &	2.90    & 217.10    &	1.90    &	        &      & $<114.3$ &  K                 & 20  \\      
 $1556-44  \tablenotemark{*}    $  & 334.50 &	6.40    & 59.00     &	2.00    &	        &      & $< 29.5$ &  K                 & 40   \\      
 $0736-40  \tablenotemark{\dag} $  & 254.20 &	-9.20   & 161.00    &	2.10    &	        &      & $< 76.7$ &  K                 & 4,9,14,43 \\      
 $1449-64  \tablenotemark{*}    $  & 315.70 &	-4.40   & 71.00     &	2.50    &	        &      & $< 28.4$ &  K                 & 40    \\      
 $2319+60                       $  & 112.10 &	-0.60   & 93.80     &	2.60    &	        &      & $< 36.1$ &  K                 & 11,12,15 \\ 
 $1323-58  \tablenotemark{*}    $  & 307.50 &  3.60     & 286.0     & 3.00      &         &      & $< 95.3  $ &  K                 & 44 \\
 $2020+28                       $  & 68.90  &	-4.70   & 24.60     &	3.10    &	        &      & $<  7.9$ &  K                 & 11,12,17 \\      
 $2016+28                       $  & 68.10  &	-4.00   & 14.20     &	3.20    &	        &      & $<  4.4$ &  K                 & 6,9,17,21 \\ 
 $1821-19  \tablenotemark{*}    $  & 12.30  & -3.10   & 224.30    & 3.20    &         &      & $<  70.1 $ &  K                 & 54 \\     
 $2255+58                       $  & 108.80 &	-0.60   & 151.10    &	3.30    &	        &      & $< 45.8$ &  K                 & 24  \\      
 $1757-24  \tablenotemark{*}    $  & 5.26   &	-0.88   & 289.00    &	3.50    &	        &      & $< 82.6$ &  G5.4-1.2 SNR      & 31    \\     
 $1703-40  \tablenotemark{*}    $  & 345.70 &	-0.20   & 360.00    &	3.80    &	        &      & $< 94.7$ &  K                 & 41   \\      
 $1648-42  \tablenotemark{*}    $  & 342.50 &	0.90    & 525.00    &	4.80    &	        &      & $<109.4$ &  K                 & 41    \\      
 $1933+16                       $  & 52.40  &	-2.10   & 158.50    &	5.20    &	        &      & $< 30.5$ &  K                 & 5,7,9,15 \\      
 $1356-60                       $  & 311.20 &	1.10    & 295.00    &	5.60    &	        &      & $< 52.7$ &  K                 & 18  \\      
 $1855+02                       $  & 35.60  &	-0.40   & 506.00    &	6.90    &	        &      & $< 73.3$ &  K                 & 21  \\     
 $1818-04                       $  & 25.50  &	4.70    & 84.40     &	        &	        & 1.60 & $> 52.8$ &  K                 & 10  \\          
 $1822-09                       $  & 21.40  &	1.30    & 19.90     &	        &	        & 1.90 & $> 10.5$ &  K                 & 11,54 \\      
 $1944+17                       $  & 55.30  &	-3.50   & 16.30     &	        &	        & 1.90 & $>  8.6$ &  K                 & 20  \\      
 $1919+21                       $  & 55.80  &	3.50    & 12.40     &	        &	        & 2.80 & $>  4.4$ &  K                 & 17  \\ 
 $1737-30  \tablenotemark{*}    $  & 358.30 & 0.20    & 153.00    &         &         & 5.50 & $>   27.8$ &  K                 & 54    \\  
 $1742-30  \tablenotemark{*}    $  & 358.60 & -1.00   & 88.80     &         &         & 5.50 & $>   16.2$ &  K                 & 54    \\  
 $0959-54  \tablenotemark{\dag} $  & 280.20 &	0.10    & 131.00    &	        &	        & 6.90 & $> 19.0$ &  K                 & 18,40 \\      
 $0940-55  \tablenotemark{*}    $  & 278.60 &	-2.20   & 180.00    &	        &         & 7.50 & $> 24.0$ &  K                 & 43    \\
 $0905-51  \tablenotemark{*}    $  & 272.2  & -3.0    &   104.00      &         &         & 8.00 & $>  13.0 $ &  K                 & 44   
\tablenotetext{*}{New pulsar distance determination since Frail \& Weisberg (1990).}
\tablenotetext{\dag}{Revised distance estimate since Frail \& Weisberg (1990).}
\tablenotetext{1}{Methods of determining the pulsar distacnes are kinematic (K), trigonometric parallax ($\Pi$), timing parallax (T), X-ray luminosity model (X), spectroscopic parallax of binary companion (SP), or association
with either supernova remnants of known distance (SNR), globular clusters, or the Small or Large Magellanic Clouds. In the cases where more than one method was used, we note in boldface which method (and reference) we chose for the tabulated distance.}
\tablerefs{References: 1. de Jager et al. (1968), 2. Gu\'elin et al. (1969), 3. Gordon, Gordon, and Shalloway (1969), 4. Manchester, Murray, and
Radhakrishnan (1969), 5. Gordon and Gordon (1970), 6. Encrenaz and Gu\'elin (1970), 7. Guelin, Encrenaz, and Bonazzola (1971), 8. G\'omez-Gonz\`alez et al
(1972), 9. Gordon and Gordon (1973), 10. G\'omez-Gonz\`alez et al. (1973), 11. G\'omez-Gonz\`alez and Guelin (1974), 12. Graham et al. (1974), 13. Gordon and
Gordon (1975), 14. Ables and Manchester (1976), 15. Booth and Lyne (1976), 16. Weisberg, Boriakoff, and Rankin (1979), 17. Weisberg, Rankin, and
Boriakoff (1980), 18. Manchester, Wellington, and McCulloch (1981), 19. Heiles et al. (1983), 20. Weisberg, Rankin, and Boriakoff (1987), 21. Clifton
et al. (1988), 22. Frail \& Weisberg (1990), 23. Kulkarni, Djorgovski, and Klemola (1991), 24. Frail et al. (1990),25. Trimble and Woltjer (1971), 26. Caswell
et al. (1975), 27. Salter, Lyne, and Anderson (1979), 28. Backer and Sramek (1982), 29. Blair et al. (1984), 30. Gwinn et al. (1986), 31. Caswell et al
(1987), 32. Feast and Walker (1987), 33. Bailes et al. (1990), 34. Ryba and Taylor (1991), 35. Caswell et al. (1992), 36. Frail, Kulkarni, and Vasisht
(1993), 37. Camilo, Foster, and Wolszczan (1994), 38. Kaspi, Taylor, and Ryba (1994), 39. Campbell (1995), 40. Koribalski et al. (1995), 41. Weisberg
et al. (1995), 42. Campbell et al. (1996), 43. Johnston et al. (1996), 44. Saravanan et al. (1996), 45. Fesen et al. (1997), 46. Sandhu et al. (1997), 47.
Cha, Sembach, and Danks (1999), 48. Fomalont et al. (1999), 49. Golden and Shearer (1999), 50. Stairs et al. (1999), 51. Toscano et al. (1999), 52.
Brisken et al. (2000), 53. Chatterjee et al. (2001), 54. Johnston et al. (2001), 55. D'Amico et al. (2001), 56. Harris (1996) and updates on
http://physun.mcmaster.ca/~harris/WEHarris.html, 57. Finley \& Ogelman (1994), 58. Legge (2000), 59. Johnston et al. (1994), 60. Cudworth and Rees (1990) , 
 61. Alcaino et al. (1987) ,
62. Alcaino (1983), 63. Buonanno et al. (1986), 64. Hesser et al. (1987), 65. Rees and Cudworth (1991), 
66. Brocato et al. (1996b), 67. Heitsch and Richtler (1999), 68. Brocato, Castellani, and Ripepi (1996a), 69. Sandquist et al. (1996), 70. Ortolani, Barbuy, and
Bica (1996), 71. Paltrinieri et al. (1998), 72. Sarajedini and Norris (1994), 73. Armandroff (1988), 74. Ortolani, Barbuy, and Bica (1994),   75. Durrell and
Harris (1993) , 76. Rey et al. (1998)  }
\enddata
\end{deluxetable}



\makeatletter
\def\jnl@aj{AJ}
\ifx\revtex@jnl\jnl@aj\let\tablebreak=\\\fi
\makeatother


\begin{deluxetable}{crrrr}
\tablecolumns{5}
\tablewidth{0pc} 
\tablecaption{Best fit parameters.}
\tablehead{
\colhead{ }      & 
\colhead{$n(R=R_{\sun},z=0) [{\rm cm^{-3}}]$}      & 
\colhead{$Z [\kpc]$}      & 
\colhead{$R [\kpc]$}  &
\colhead{$A$}
}
\startdata
$\sech^2(x)$ & &&&\\
             & $ 1.77 \times 10^{-2} $ &  1.10  &  15.4  &  0.30  \\
             & $ 1.07 \times 10^{-2} $ &  0.04 &    3.6 &        \\
$\exp(-x)$   & &&&\\
             & $ 2.03 \times 10^{-2} $ &  1.07  &  30.4 &  0.31  \\
             & $ 0.71 \times 10^{-2} $ &  0.05 &    1.5 &        \\
\enddata
\end{deluxetable}



\makeatletter
\def\jnl@aj{AJ}
\ifx\revtex@jnl\jnl@aj\let\tablebreak=\\\fi
\makeatother


\begin{deluxetable}{rrrrrrrrl}
\tablecolumns{9}
\tablecaption{Outlier pulsars}
\tablehead{\colhead{PSR} &  
\colhead{l}      & 
\colhead{b}      & 
\colhead{D\tablenotemark{a}}      &
\colhead{$DM_{data}$}  &
\colhead{$DM_{-}\tablenotemark{b}$} &
\colhead{$DM_{+}\tablenotemark{b}$} &
\colhead{$DM_{excess}$} &
\colhead{Method\tablenotemark{c}} 
}

\startdata
$0656+14$  &  201.11 &    8.26 & 0.18 &   14.0 &    3.3  &    6.2&   7.8 &  X \\
$0833-45$  &  263.55 &   -2.79 & 0.25 &   68.2 &    4.9  &    9.0&  59.2 &  Vela \\
$1741-11$  &   14.79 &    9.18 & 0.36 &    3.1 &    6.3  &   11.7&  -3.2 &  TP \\
$0823+26$  &  197.00 &   31.70 & 0.36 &   19.5 &    4.9  &    9.1&  10.4 &  $\Pi$\\
$1259-63$  &  304.20 &   -0.99 & 1.10 &  146.7 &   22.8  &   42.4& 104.3 &  SP\\
$1807-24$  &    5.80 &   -2.20 & 2.50 &  134.0 &   44.8  &   83.2&  50.8 &  NGC 6544\\
$1054-62$  &  290.30 &   -3.00 & 2.70 &  321.0 &   40.2  &   74.6& 246.4 &  K\\
$1900+01$  &   35.70 &   -2.00 & 3.40 &  246.4 &   57.5  &  106.8& 139.6 &  K\\
$1859+07$  &   40.60 &    1.10 & 3.75 &  261.0 &   75.1  &  139.4& 121.6 &  K\\
$1509-58$  &  320.32 &   -1.16 & 4.40 &  253.2 &   84.3  &  156.6&  96.6 &  MSH 15-52 SNR\\
$1641-45$  &  339.20 &   -0.20 & 4.60 &  475.0 &  179.2  &  332.8& 142.2 &  K\\
$1718-35$  &  351.70 &    0.70 & 4.80 &  496.0 &  137.9  &  256.0& 240.0 &  K\\
$1758-23$  &    6.80 &   -0.10 & 5.20 & 1074.0 &  266.9  &  495.6& 578.4 &  K, W28\\
$1338-62$  &  308.73 &   -0.04 & 6.90 &  730.0 &  204.5  &  379.8& 350.2 &  G308.8-0.1 SNR\\
$1937+21$  &   57.51 &   -0.29 & 9.70 &   71.0 &  207.6  &  385.5&-136.6 &  K,TP\\
\enddata
\tablenotetext{a}{For pulsars with kinematic distance, $D=0.5(D_{min}+D_{max})$.}
\tablenotetext{b}{$DM_{\pm}=(1 \pm A) DM_{model}(D)$}
\tablenotetext{c}{See Table 1 for list of methods.}
\end{deluxetable}


\end{document}